\documentclass[11pt]{article}

\usepackage[preprint]{acl}

\usepackage{times}
\usepackage{latexsym}

\usepackage[T1]{fontenc}
\usepackage[utf8]{inputenc}

\usepackage{microtype}

\usepackage{inconsolata}

\usepackage{graphicx}
\usepackage{amsmath}
\usepackage{amsfonts}
\usepackage{booktabs}
\usepackage{multirow}

\usepackage{xcolor}
\usepackage{colortbl}
\usepackage{booktabs}
\usepackage{array}
\usepackage{multirow}

\usepackage{graphicx}  % 用于 \scalebox 和 \raisebox
\usepackage{twemojis}
\usepackage{booktabs,threeparttable}

\title{ 
Do Models See in Line with Human Vision?
Probing the Correspondence Between LVLM Representations and EEG Signals}

\author{%
\textbf{Xin Xiao$^{1}$}, \textbf{Yang Lei$^{1}$}, \textbf{Haoyang Zeng$^{1}$}, \textbf{Xiao Sun$^{1}$},\\
\textbf{Xinyi Jiang$^{2}$}, \textbf{Yu Tian$^{3}$}, \textbf{Hao Wu$^{4}$}, \textbf{Kaiwen Wei$^{1}$}, \textbf{Jiang Zhong$^{1}$} \\
\\
$^{1}$Chongqing University \quad $^{2}$UNSW Sydney \quad $^{3}$Tsinghua University \\
$^{4}$The First Affiliated Hospital of Chongqing Medical University \\
}

\begin{document}
\maketitle
\begin{abstract}
Large Vision Language Models (LVLMs) exhibit strong visual understanding and reasoning abilities. However, whether their internal representations reflect human visual cognition is still under-explored. In this paper, we address this by quantifying LVLM–brain alignment using image-evoked Electroencephalogram (EEG) signals, analyzing the effects of model architecture, scale, and image type.  Specifically, by using ridge regression and representational similarity analysis, we compare visual representations from 32 open-source LVLMs with corresponding EEG responses.
We observe a structured LVLM–brain correspondence: First, intermediate layers (8–16) show peak alignment with EEG activity in the 100–300 ms window, consistent with hierarchical human visual processing. Secondly, multimodal architectural design contributes 3.4× more to brain alignment than parameter scaling, and models with stronger downstream visual performance exhibit higher EEG similarity. Thirdly, spatiotemporal patterns further align with known cortical visual pathways.  These results demonstrate that LVLMs learn human-aligned visual representations and establish neural alignment as a biologically grounded benchmark for evaluating and improving LVLMs. 
In addition, those results could provide insights that may inform the development of neuro-inspired applications.

\end{abstract}

\section{Introduction}
Large Vision Language Models (LVLMs) learn unified visual–semantic representations for image captioning, visual question answering, and image–text retrieval by mapping visual and linguistic inputs into a shared high-dimensional space \cite{yin2024survey, fei2024multimodal, li2024seed, yang2025thinking, yang2025mage}. 
In neuroscience, neuroimaging methods provide rich measurements of the spatiotemporal dynamics of brain activity during visual cognition \cite{cichy2020eeg}, capturing neural responses critical for studying human visual perception \cite{mustafa2012eeg}. This motivates investigating the similarity between LVLM representations and image-evoked neural signals. Such investigation may ultimately inform neuroscience-grounded, human-aligned multimodal AI systems.

\begin{figure}[t]
  \includegraphics[width=\columnwidth]{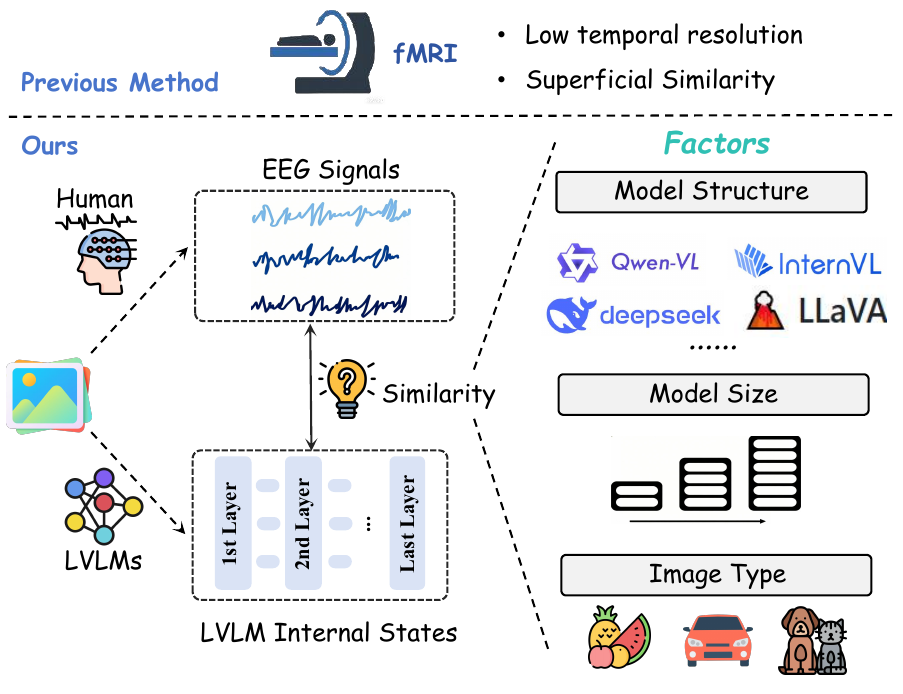}
  \caption{The motivation of this paper: exploring the similarity of human brain EEG signals and LVLM  representations from various aspects.}
  \label{fig:introduction}
\end{figure}

Previous studies show strong alignment between deep learning models and brain cognitive signals. CNNs predict activity along the primate ventral visual stream, with deeper layers corresponding to higher-level cortical areas \cite{yamins2014performance, kazemian2025convolutional}, while Vision Transformers suggest that self-attention captures long-range dependencies relevant to biological vision \cite{tuli2021convolutional, raghu2021vision, hernandez2025vision}. Beyond discriminative models, generative approaches such as latent diffusion enable visual decoding from brain signals, revealing semantically aligned neural representations \cite{wang2024mindbridge, jing2025beyond}. Overall, models from traditional deep learning architectures to LVLMs exhibit consistent structural and functional correspondences with human brain processing and can reliably predict and decode brain activity \cite{mischler2024contextual, doerig2025high, oota2025correlating}.

% Previous work has shown strong alignment between deep learning models and brain cognitive signals. CNNs effectively predict primate ventral visual stream activity, with deeper layers corresponding to higher-level cortical areas \cite{yamins2014performance,kazemian2025convolutional}, while Vision Transformers suggest that self-attention captures complementary long-range dependencies relevant to biological vision \cite{tuli2021convolutional, raghu2021vision, hernandez2025vision}.
% In parallel, generative models such as latent diffusion enable brain decoding from fMRI and EEG, reconstructing visual stimuli and revealing semantically aligned neural representations \cite{wang2024mindbridge,jing2025beyond}. More broadly, recent studies demonstrate that deep learning models, from LLMs to LVLMs, exhibit structural and functional correspondences with human brain information processing and can reliably predict and decode brain activity \cite{mischler2024contextual, doerig2025high, oota2025correlating}.

However, existing similarity analysis research primarily relies on functional Magnetic Resonance Imaging (fMRI) data \cite{logothetis2008we}, which offers high spatial resolution but suffers from limited temporal resolution, as shown in Figure~\ref{fig:introduction}. In contrast, EEG signals \cite{cichy2020eeg, mustafa2012eeg} provide millisecond-scale temporal precision that captures the dynamic flow of cognitive processing, yet their alignment with existing LVLMs remains underexplored. More importantly, it remains unclear whether any observed alignment between LVLMs and brain signals reflects superficial representational similarity or deeper computational parallels. Human visual cognition is inherently dynamic and hierarchical, with distinct processing stages unfolding over time and across cortical regions \cite{turner2023visual, kaas2010imagery}. Whether LVLMs exhibit analogous temporal organization when encoding visual stimuli, and which factors contribute to such alignment, are still largely unexplored.

% However, existing similarity analysis research primarily relies on fMRI data, which offers high spatial resolution but suffers from limited temporal resolution. In contrast, EEG signals provide millisecond-scale temporal precision that captures the dynamic flow of cognitive processing, yet their alignment with state-of-the-art LVLMs remains unexplored.

% However, it remains unclear whether any observed alignment between LVLMs and brain signals reflects superficial representational similarity or deeper computational parallels. Human visual cognition is inherently dynamic and hierarchical, with distinct processing stages unfolding over time and across cortical regions. Whether LVLMs exhibit analogous temporal organization when encoding visual stimuli, and which factors contribute to such alignment, are still largely unexplored. 

Motivated by these considerations, this paper investigates the extent to which LVLMs reflect image-induced EEG signals and analyzes how different factors, including model architecture, model scale, and image type, influence LVLM–brain alignment.  
Specifically, we employ ridge regression \cite{mcdonald2009ridge} and Representational Similarity Analysis (RSA) \cite{kriegeskorte2008representational}, two methods widely used in neuroscience to compare neural and model representations \cite{mischler2024contextual, zhou2024divergences, doerig2025high}, to quantify the similarity between LVLM multimodal representations and image-evoked EEG signals.
We systematically analyze LVLM–brain alignment using 32 open-source models spanning architectures and scales, revealing that multimodal architectural design contributes 3.4× more to brain alignment than parameter scaling. Through hierarchical layer–time analysis, we find intermediate layers (8–16) align best with EEG activity during 100–300 ms, matching the timeline of human visual processing. Stronger LVLM task performance correlates with higher EEG similarity, and spatial alignment follows established cortical pathways. 
Our main contributions can be summarized as follows:

% \begin{enumerate}

(1) To the best of our knowledge, this is the first work to explore LVLM–EEG alignment. We leverage image-evoked EEG signals with ridge regression and representational similarity analysis to compare LVLM representations with human brain responses across 32 models, enabling systematic analysis of architectural and scaling effects. 
% We present the first EEG-based analysis using image-evoked signals with ridge regression and RSA to quantify similarity between LVLM representations and human brain responses across 32 models, enabling systematic analysis of scale and architecture effects.
% We use ridge regression and RSA to quantify the similarity between LVLM visual representations and image-evoked brain responses across 32 representative LVLMs, analyzing the effects of model scale and architectural design.

(2) We find that LVLMs show strong alignment with human visual processing from both representational hierarchy and temporal dynamics. Multimodally trained and larger models outperform vision-only models, with intermediate layers exhibiting peak alignment during feature integration.

(3) We find that the spatiotemporal alignment mirrors known cortical visual pathways and, from the perspective of task performance, correlates strongly with downstream visual benchmarks, indicating that LVLM–brain similarity is a meaningful measure of human-aligned visual understanding.

    % \item we employ ridge regression and RSA to quantitatively evaluate the similarity between LVLMs and human visual cognitive processing. By analyzing 32 representative LVLMs, we systematically investigate how factors such as model scale and architectural design influence the alignment between LVLM visual representations and brain responses elicited by images.

    % \item Multimodal language models demonstrate significant alignment with human visual processing, both in representational hierarchy and temporal dynamics.  Larger models with multimodal training exhibit stronger neural correspondence than vision-only architectures, while intermediate model layers show peak alignment during the critical time window associated with object feature integration in human vision. 

    % \item The spatiotemporal progression of this alignment closely mirrors established cortical visual processing pathways, indicating that LVLMs implicitly learn biologically plausible visual-semantic transformations. There is a strong consistency between LVLM performance on downstream visual tasks and their similarity to image-induced brain signals. This finding indicates that LVLM–brain visual similarity serves as a meaningful indicator for evaluating the human-aligned visual understanding capabilities of LVLMs.
% \end{enumerate}

    % We show that LVLM–brain correspondence goes beyond static similarity and reflects consistent visual processing pathways, with different LVLM stages aligning with distinct temporal phases of EEG activity, indicating parallels in hierarchical visual processing.

\section{Related Work}

\subsection{Neural Foundations of Vision Processing}
Human visual processing involves functionally specialized cerebral lobes, and EEG captures its spatiotemporal dynamics across processing stages. Visual processing originates in the occipital lobe, encompassing primary and secondary visual cortices V1–V4 that encode low-level features such as contrast, edges, orientation, and spatial location \cite{turner2023visual, kaas2010imagery}, with occipital EEG responses emerging around 80–100 ms after stimulus onset \cite{robinson2017very}. 
Visual information then propagates along the ventral "what" pathway in the temporal lobe for object recognition and semantic processing \cite{murray2007visual, kaas2010imagery}, and the dorsal "where/how" pathway in the parietal lobe for spatial encoding and visuomotor integration \cite{babiloni2006visuo, nassi2009parallel}; these high-level visual cognitive processes are manifested in EEG signals as sustained neural activity at 200–400 ms, with semantic-related visual image processing peaking at approximately 300 ms \cite{grutzner2010neuroelectromagnetic}. The spatiotemporal latency characteristics of EEG responses precisely reflect the stepwise transformation of low-level visual features into high-level meaningful visual representations in human brain.

\begin{figure*}[t]
  \includegraphics[width=\textwidth]{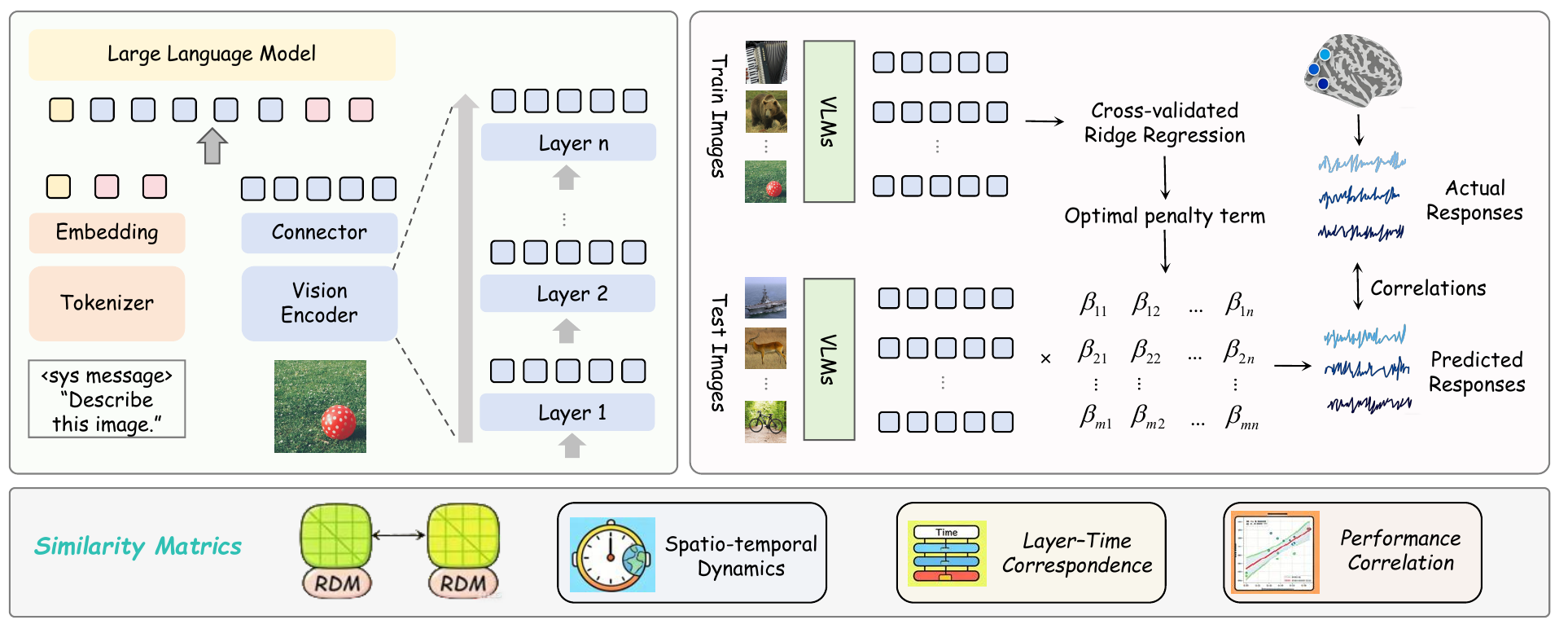}
    % \caption{Overview of the method for investigating processing similarities between the human brain EEG signals and LVLMs.  Using identical visual stimuli, image representations are extracted from different layers of LVLMs and mapped to human brain responses via ridge regression to quantify model–brain representational alignment. }
    \caption{Illustration of the analysis pipeline for exploring representational similarity between human EEG signals and LVLMs. Image representations from different LVLM layers are compared with EEG responses elicited by the same visual stimuli to assess model--brain alignment.}
  \label{fig:framework}
\end{figure*}

\subsection{LVLM-Brain Alignment}
Emerging research reveals strong parallels between LVLMs and human brain representations, with most evidence to date derived from fMRI-based neuroimaging. Neuroimaging studies show that high-level visual brain representations align with LLM embeddings, as scene-caption embeddings can model neural responses to natural scenes and enable caption reconstruction from brain signals \cite{doerig2025high}. LVLMs also encode human-like object concepts corresponding to category-selective regions such as the fusiform face area and parahippocampal place area, and multimodal encoding models demonstrate that LVLM-derived representations outperform unimodal features in predicting fMRI responses \cite{du2025human, oota2025multi}. In addition, instruction-tuned LVLMs exhibit enhanced alignment with the brain’s vision–language integration mechanisms as measured with fMRI \cite{oota2025correlating}. Complementary evidence from systematically varied self-supervised vision transformers (DINOv3) further shows that model size, training data scale, and image type jointly shape brain similarity when compared against ultra-high-field fMRI and MEG recordings \cite{raugel2025disentangling}. However, most existing evidence relies on fMRI and lacks millisecond-level analysis of LVLM–brain temporal dynamics, motivating us to conduct the EEG-LVLM similarity investigation.

% Emerging research reveals strong parallels between LVLMs and fMRI in representational processing. Neuroimaging studies show that high-level visual brain representations align with LLM embeddings, as scene-caption embeddings can model neural responses to natural scenes and enable caption reconstruction from brain signals \cite{doerig2025high}. LVLMs also encode human-like object concepts corresponding to category-selective regions such as FFA and PPA, and multimodal encoding models demonstrate that LVLM-derived representations outperform unimodal features in predicting brain responses \cite{du2025human, oota2025multi}. In addition, instruction-tuned LVLMs exhibit enhanced alignment with the brain’s vision–language integration mechanisms \cite{oota2025correlating}. Complementary evidence from systematically varied self-supervised vision transformers (DINOv3) shows that model size, training data scale, and image type jointly shape brain similarity when compared against ultra-high field fMRI and MEG recordings \cite{raugel2025disentangling}.

\section{Methodology}
\label{method}
To evaluate the similarity between LVLMs and human visual processing, as shown in Figure \ref{fig:framework}, we use ridge regression \cite{mcdonald2009ridge} to linearly map LVLM image features to EEG-derived brain signals elicited by the same visual stimuli, thereby quantifying LVLM–brain representational alignment. Based on this framework, we examine how model architecture, scale, and image category affect LVLM–brain similarity, analyze spatiotemporal correspondences with human visual processing, and explore the relationship between LVLM–brain alignment and LVLM performance on vision-related benchmarks.

% To evaluate the similarity between LVLMs and human visual processing, we adopt ridge regression to construct a linear mapping for comparing representational similarity between LVLMs and the human brain.  Specifically, using identical images as visual stimuli, we fit human EEG-derived brain activation signals with image features extracted from LVLMs, thereby quantifying the degree of alignment between model and neural representations. as shown in Figure \ref{fig:framework}.  
% Based on this framework, we investigate the effects of model architecture, scale, and image category on LVLM–brain representational similarity, and further analyze the spatiotemporal correspondence between LVLMs and the human visual processing, and explore the intrinsic relationship between LVLM–brain alignment and LVLM performance on vision-related benchmark tasks.

\subsection{Brain Data and Preprocessing}
This study focuses on a publicly available large-scale within-subject EEG dataset, the THINGS-EEG dataset \cite{gifford2022large}, which contains EEG recordings from 10 subjects collected under the Rapid Serial Visual Presentation (RSVP) paradigm \cite{grootswagers2019representational}. 
The training set comprises 1,654 object concepts with 10 images each, repeated four times per subject, while the test set includes 200 concepts with a single image repeated 80 times per subject. Following prior work \cite{li2024visual}, EEG data were band-pass filtered between 0.1 and 100 Hz, segmented from 0 to 1000 ms after stimulus onset with baseline correction using the 200 ms pre-stimulus interval, and z-score normalized together with image features. Principal component analysis (PCA) \cite{abdi2010principal} was then applied for dimensionality reduction prior to regression-based alignment and representational similarity analyses.

% \subsection{Details on the LVLMs}
\subsection{Extraction of Image Embeddings}
% We used 32 LVLMs from nine different model families, covering both vision-only and vision--language architectures with varying model scales, including ViT-based models \cite{dosovitskiy2020image}, Qwen2.5-VL \cite{bai2025qwen2}, Qwen3-VL \cite{bai2025qwen3vltechnicalreport}, LLaVA-v1.5 \cite{liu2024improved}, LLaVA-NeXT \cite{liu2024llavanext}, InternVL3 \cite{zhu2025internvl3exploringadvancedtraining}, InternVL3.5 \cite{wang2025internvl35advancingopensourcemultimodal}, DeepSeek-VL2 \cite{wu2024deepseekvl2mixtureofexpertsvisionlanguagemodels}, and SAIL-VL2 \cite{yin2025sailvl2technicalreport},
% % including ViT-based models \cite{dosovitskiy2020image}, Qwen-VL series, LLaVA variants, InternVL, DeepSeek-VL, and SAIL-VL. 
% All models employed in this study are publicly available from \url{https://huggingface.co/models}. 
% For each LVLMs, image representations were obtained by averaging the embeddings of all visual tokens from the final layer of the vision encoder (or vision tower). In addition, we extracted layer-wise visual representations to analyze the correspondence between different model layers and human brain responses. These extracted image representations were then used as inputs for subsequent regression-based alignment and representational similarity analyses.

We extracted visual features using LVLMs by averaging the embeddings of all visual tokens from the final layer of the vision encoder. Additionally, we obtained layer-wise visual representations to analyze the correspondence between different model layers and human brain responses. These features were then used as inputs for subsequent regression-based alignment and representational similarity analyses.

\subsection{Mapping Image Embeddings to EEG Responses}
Following prior work \cite{mischler2024contextual, zhou2024divergences, doerig2025high}, we use ridge regression \cite{mcdonald2009ridge} to quantify the correspondence between LVLM visual representations and EEG responses. We evaluate how well image embeddings from different model layers predict EEG activity at individual sensor channels.
Let $N$ be the number of stimulus samples. Neural responses are collected in $\mathbf{y} \in \mathbb{R}^{N \times D}$. LVLM visual representations are denoted by $\mathbf{X} \in \mathbb{R}^{N \times L \times D}$, where $L$ is the number of layers and $D$ the embedding dimension. For each layer $l$, we extract $\mathbf{X}^{(l)} \in \mathbb{R}^{N \times D}$.
Predictive performance is assessed using $K$-fold cross-validation. 
% In each fold $k$, ridge regression is trained on the training split and evaluated on the held-out test split. 
The prediction accuracy for layer $l$ is defined as the average Pearson correlation across folds:
\begin{equation}
\rho_l = \frac{1}{K} \sum_{k=1}^{K}
\mathrm{corr}\!\left(
\mathbf{y}_{\mathrm{test}}^{(k)},
\mathbf{X}_{\mathrm{test}}^{(l,k)} \hat{\boldsymbol{\beta}}^{(l,k)}
\right),
\end{equation}
where $\mathbf{X}_{\mathrm{test}}^{(l,k)} \in \mathbb{R}^{N_k \times D}$ and $\mathbf{y}_{\mathrm{test}}^{(k)} \in \mathbb{R}^{N_k \times D}$ denote test features and EEG responses, and $\hat{\boldsymbol{\beta}}^{(l,k)} $ are regression weights.
The ridge solution is: 
\begin{equation}
\hat{\boldsymbol{\beta}}^{(l,k)} =
\left(
\mathbf{X}_{\mathrm{train}}^{(l,k)\top}
\mathbf{X}_{\mathrm{train}}^{(l,k)}
+ \alpha \mathbf{I}
\right)^{-1}
\mathbf{X}_{\mathrm{train}}^{(l,k)\top}
\mathbf{y}_{\mathrm{train}}^{(k)},
\end{equation}
where $\alpha$ is the regularization coefficient and $\mathbf{I}$ the identity matrix.

\subsection{Multidimensional Similarity Assessment}
We present our analysis from four complementary perspectives: (1) Predictive performance evaluation, (2) Spatio-temporal pattern of predictions, (3) Hierarchical LVLM-to-EEG temporal alignment, and (4) Category-dependent similarity.
% \paragraph{Predictive Performance Evaluation}
% We evaluated similarity between model-predicted and neural representations using both signal- and representation-level metrics. At the signal level, Pearson correlation measured linear correspondence between true EEG $\mathbf{y}$ and predictions $\hat{\mathbf{y}}$, while Spearman correlation captured monotonic relationships. At the representation level, centered kernel alignment (CKA) \cite{lu2014multiple} with a linear kernel compared predicted and observed EEG features.
% Given two representation matrices $\mathbf{K}$ and $\mathbf{L}$, the CKA score is defined as:
% \begin{equation}
% \mathrm{CKA}(\mathbf{K}, \mathbf{L}) = 
% \frac{\langle \mathbf{K}_c, \mathbf{L}_c \rangle_F}
% {\|\mathbf{K}_c\|_F \|\mathbf{L}_c\|_F},
% \end{equation}
% where $\mathbf{K}_c$ and $\mathbf{L}_c$ denote the centered kernel matrices, $\langle \cdot, \cdot \rangle_F$ is the Frobenius inner product, and $\|\cdot\|_F$ is the Frobenius norm. Representational similarity analysis (RSA) \cite{kriegeskorte2008representational} used correlation-based dissimilarity matrices, with alignment quantified via Spearman correlation (RSA score) and Kendall’s $\tau$ \cite{noether1981kendall} for rank-based robustness.

\paragraph{Predictive Performance Evaluation.}
We evaluated the similarity between model-predicted and neural representations using both signal- and representation-level metrics. At the signal level, the \textbf{Pearson} correlation \cite{pearson1895vii} measured the linear correspondence between the true EEG signal vector $\mathbf{y}$ and the predicted signal vector $\hat{\mathbf{y}}$, while the \textbf{Spearman} rank correlation ($\rho$) \cite{spearman1904} captured monotonic relationships. At the representation level, we employed \textbf{centered kernel alignment (CKA)} \cite{cortes2012algorithms} with a linear kernel to compare the overall representational geometry of the predicted and observed EEG features.
Formally, let $\mathbf{K}$ and $\mathbf{L}$ be the similarity (kernel) matrices derived from the predicted EEG features and the true EEG features across all stimulus trials, respectively. The linear CKA index is defined as:
\begin{equation}
\mathrm{CKA}(\mathbf{K}, \mathbf{L}) = 
\frac{\langle \mathbf{K}_c, \mathbf{L}_c \rangle_F}
{\|\mathbf{K}_c\|_F \|\mathbf{L}_c\|_F},
\end{equation}
where $\mathbf{K}_c$ and $\mathbf{L}_c$ denote the centered versions of $\mathbf{K}$ and $\mathbf{L}$, $\langle \cdot, \cdot \rangle_F$ is the Frobenius inner product, and $\|\cdot\|_F$ is the Frobenius norm. Additionally, \textbf{representational similarity analysis (RSA) } \cite{kriegeskorte2008representational} was performed by constructing correlation-based representational dissimilarity matrices (RDMs) for both the predicted and neural representations. The alignment between these two RDMs was quantified using the Spearman correlation (RSA score) and \textbf{Kendall}’s rank correlation coefficient ($\tau$) \cite{kendall1938new} to ensure robustness to non-linear monotonic relationships.

\paragraph{Spatio-temporal Pattern of Predictions.}
To examine the spatial relationship between LVLM  features and EEG signals, ridge regression was used to predict each EEG channel from LVLM features, and Pearson correlations between predicted and true signals were computed. Channel-level correlations were visualized on a standard 10-20 EEG montage using the MNE toolbox \cite{gramfort2013meg}. Channels were further grouped into 4 functional regions: Frontal, Central, Parietal, and Occipital, and regional mean and standard deviation of correlations quantified predictive performance.

\paragraph{Layer-wise LVLM-EEG Temporal Alignment.}
% To examine temporal alignment between LVLM layer representations and EEG dynamics, we conducted a cross-dimensional similarity analysis. EEG signals were segmented into 100 ms windows and flattened, while features were extracted from each LVLM layer. After standardization and dimensionality reduction, ridge regression was applied to model the correspondence between EEG activity and layer-wise LVLM representations across time.
To examine temporal alignment between LVLM layer representations and EEG dynamics, we conducted a cross-dimensional similarity analysis. EEG signals were segmented into 100,ms windows and flattened, while features were extracted from each LVLM layer, then following the ridge regression procedure described in Section~3.3.
% After standardization and dimensionality reduction, the correspondence between EEG activity and layer-wise LVLM representations was estimated following the ridge regression procedure described in Section~3.3.

\begin{table*}[ht]
\centering
\resizebox{0.88\textwidth}{!}{
\begin{tabular}{clccccc}
\toprule
% \rowcolor{gray!10}
\textbf{Model Series} & \textbf{Scale} & \textbf{Pearson} & \textbf{Spearman} & \textbf{CKA} & \textbf{RSA} & \textbf{Kendall} \\
\midrule

\multirow{3}{*}{ViT}
& B & \cellcolor{teal!18}{0.2262 ± 0.0314} & \cellcolor{teal!18}{0.2187 ± 0.0289} & \cellcolor{teal!18}{0.4033 ± 0.0923} & \cellcolor{teal!18}{0.2803 ± 0.0923} & \cellcolor{teal!18}{0.1492 ± 0.0203} \\
& L & 0.2168 ± 0.0298 & 0.2106 ± 0.0284 & 0.4001 ± 0.0920 & 0.2800 ± 0.0856 & 0.1434 ± 0.0197 \\
& H & 0.2152 ± 0.0315 & 0.2083 ± 0.0291 & 0.3878 ± 0.0905 & 0.2632 ± 0.0824 & 0.1420 ± 0.0203 \\
\midrule

\multirow{4}{*}{Qwen2.5-VL}
& 3B & \cellcolor{teal!24}{0.2427 ± 0.0324} & \cellcolor{teal!24}{0.2361 ± 0.0294} & 0.4135 ± 0.0950 & \cellcolor{teal!24}{0.2976 ± 0.0866} & \cellcolor{teal!24}{0.1614 ± 0.0207} \\
& 7B & 0.2316 ± 0.0280 & 0.2247 ± 0.0262 & 0.4151 ± 0.0970 & \cellcolor{teal!24}{0.2976 ± 0.0879} & 0.1535 ± 0.0183 \\
& 32B & 0.2370 ± 0.0288 & 0.2301 ± 0.0264 & 0.4135 ± 0.0954 & 0.2953 ± 0.0900 & 0.1572 ± 0.0185 \\
& 72B & 0.2343 ± 0.0294 & 0.2269 ± 0.0280 & \cellcolor{teal!24}{0.4166 ± 0.0997} & 0.2940 ± 0.0904 & 0.1550 ± 0.0196 \\
\midrule

\multirow{4}{*}{Qwen3-VL}
& 2B & 0.2321 ± 0.0305 & 0.2245 ± 0.0277 & 0.4083 ± 0.0925 & 0.2883 ± 0.0881 & 0.1532 ± 0.0195 \\
& 4B & 0.2379 ± 0.0315 & 0.2306 ± 0.0284 & 0.4196 ± 0.0966 & 0.3056 ± 0.0929 & 0.1576 ± 0.0200 \\
& 8B & \cellcolor{teal!30}{0.2385 ± 0.0314} & \cellcolor{teal!30}{0.2310 ± 0.0288} & \cellcolor{teal!30}{0.4250 ± 0.1005} & \cellcolor{teal!30}{0.3138 ± 0.0917} & \cellcolor{teal!30}{0.1578 ± 0.0202} \\
& 32B & 0.2338 ± 0.0316 & 0.2263 ± 0.0292 & 0.4183 ± 0.0985 & 0.2996 ± 0.0889 & 0.1547 ± 0.0204 \\
\midrule

\multirow{2}{*}{LLaVA-v1.5}
& 7B & \cellcolor{teal!12}{0.2102 ± 0.0297} & \cellcolor{teal!12}{0.2050 ± 0.0282} & \cellcolor{teal!12}{0.3815 ± 0.0901} & \cellcolor{teal!12}{0.2653 ± 0.0846} & 0.1395 ± 0.0197 \\
& 13B & \cellcolor{teal!12}{0.2102 ± 0.0297} & \cellcolor{teal!12}{0.2050 ± 0.0282} & 0.3814 ± 0.0902 & 0.2651 ± 0.0846 & \cellcolor{teal!12}{0.1396 ± 0.0197} \\
\midrule

\multirow{2}{*}{LLaVA-Next}
& 7B & \cellcolor{teal!36}{0.2377 ± 0.0367} & \cellcolor{teal!36}{0.2316 ± 0.0341} & \cellcolor{teal!36}{0.4267 ± 0.1002} & \cellcolor{teal!36}{0.3223 ± 0.0926} & \cellcolor{teal!36}{0.1583 ± 0.0238} \\
& 13B & 0.2358 ± 0.0355 & 0.2298 ± 0.0325 & 0.4251 ± 0.1028 & 0.3210 ± 0.0938 & 0.1570 ± 0.0228 \\
\midrule

\multirow{6}{*}{InternVL3}
& 1B & 0.2431 ± 0.0323 & 0.2357 ± 0.0302 & 0.4208 ± 0.0991 & 0.3141 ± 0.0958 & 0.1612 ± 0.0213 \\
& 2B & 0.2469 ± 0.0328 & 0.2398 ± 0.0312 & 0.4240 ± 0.0990 & 0.3182 ± 0.0948 & 0.1640 ± 0.0219 \\
& 8B & 0.2521 ± 0.0341 & 0.2437 ± 0.0324 & 0.4340 ± 0.1027 & 0.3297 ± 0.0970 & 0.1669 ± 0.0229 \\
& 9B & 0.2502 ± 0.0340 & 0.2427 ± 0.0320 & 0.4366 ± 0.1056 & 0.3340 ± 0.0986 & 0.1661 ± 0.0226 \\
& 14B & 0.2514 ± 0.0353 & 0.2429 ± 0.0332 & \cellcolor{teal!45}{0.4381 ± 0.1050} & 0.3283 ± 0.0992 & 0.1663 ± 0.0235 \\
& 38B & \cellcolor{teal!45}{0.2591 ± 0.0375} & \cellcolor{teal!45}{0.2523 ± 0.0353} & 0.4352 ± 0.1084 & \cellcolor{teal!45}{0.3339 ± 0.0985} & \cellcolor{teal!45}{0.1730 ± 0.0250} \\
\midrule

\multirow{6}{*}{InternVL3.5}
& 1B & 0.2529 ± 0.0333 & 0.2454 ± 0.0315 & 0.4316 ± 0.1008 & 0.3274 ± 0.0955 & 0.1679 ± 0.0223 \\
& 2B & 0.2454 ± 0.0343 & 0.2385 ± 0.0327 & 0.4166 ± 0.0996 & 0.3139 ± 0.0934 & 0.1632 ± 0.0230 \\
& 4B & 0.2512 ± 0.0351 & 0.2443 ± 0.0331 & 0.4281 ± 0.1013 & 0.3185 ± 0.0912 & 0.1672 ± 0.0234 \\
& 8B & 0.2560 ± 0.0356 & 0.2491 ± 0.0332 & 0.4352 ± 0.1054 & 0.3314 ± 0.0962 & 0.1706 ± 0.0235 \\
& 14B & 0.2523 ± 0.0358 & 0.2447 ± 0.0336 & 0.4371 ± 0.1042 & 0.3271 ± 0.0966 & 0.1677 ± 0.0238 \\
& 38B & \cellcolor{yellow!45}{0.2649 ± 0.0373} & \cellcolor{yellow!45}{0.2573 ± 0.0348} & \cellcolor{yellow!45}{0.4432 ± 0.1081} & \cellcolor{yellow!45}{0.3422 ± 0.0968} & \cellcolor{yellow!45}{0.1764 ± 0.0247} \\
\midrule

\multirow{3}{*}{DeepSeek-VL2}
& Tiny & 0.2368 ± 0.0332 & 0.2301 ± 0.0314 & 0.4174 ± 0.0990 & 0.3042 ± 0.0914 & 0.1572 ± 0.0220 \\
& Small & \cellcolor{teal!42}{0.2501 ± 0.0327} & \cellcolor{teal!42}{0.2428 ± 0.0303} & \cellcolor{teal!42}{0.4369 ± 0.1043} & \cellcolor{teal!42}{0.3316 ± 0.0979} & \cellcolor{teal!42}{0.1661 ± 0.0214} \\
& Base & 0.2481 ± 0.0328 & 0.2409 ± 0.0303 & 0.4350 ± 0.1058 & 0.3264 ± 0.0957 & 0.1648 ± 0.0214 \\
\midrule

\multirow{2}{*}{SAIL-VL2}
& 2B & 0.2298 ± 0.0321 & 0.2233 ± 0.0302 & 0.3974 ± 0.0954 & 0.2905 ± 0.0880 & 0.1523 ± 0.0211 \\
& 8B & \cellcolor{teal!24}{0.2370 ± 0.0332} & \cellcolor{teal!24}{0.2296 ± 0.0314} & \cellcolor{teal!24}{0.4052 ± 0.0979} & \cellcolor{teal!24}{0.3011 ± 0.0939} & \cellcolor{teal!24}{0.1567 ± 0.0220} \\

\bottomrule
\end{tabular}}
\caption{Similarity metrics between LVLMs representations and EEG signals across different model families and scales. Cells with \colorbox{teal!45}{teal background} indicate the maximum value for each metric within a single model family. The overall optimal result across all models is highlighted with \colorbox{yellow!45}{yellow background}. All values are reported as mean ± standard deviation across subjects, with all permutation-test $p$-values $< 0.05$.}
% All values are reported as mean ± standard deviation across subjects. All corresponding permutation-test $p$-values are below 0.05.}
\label{tab:vlm_eeg_similarity}
\end{table*}

\paragraph{Category-dependent Similarity.}
To examine category-dependent alignment between LVLM features and EEG responses, stimuli were grouped into 12 high-level ImageNet categories \cite{deng2009imagenet}: mammal, bird, fish, reptile, amphibian, vehicle, furniture, musical instrument, geological formation, tool, flower, and fruit.  Category labels were assigned via a zero-shot classification using bart-large-mnli \cite{lewis2019bartdenoisingsequencetosequencepretraining}

\begin{figure*}[t]
  \centering
  \includegraphics[width=0.9\textwidth]{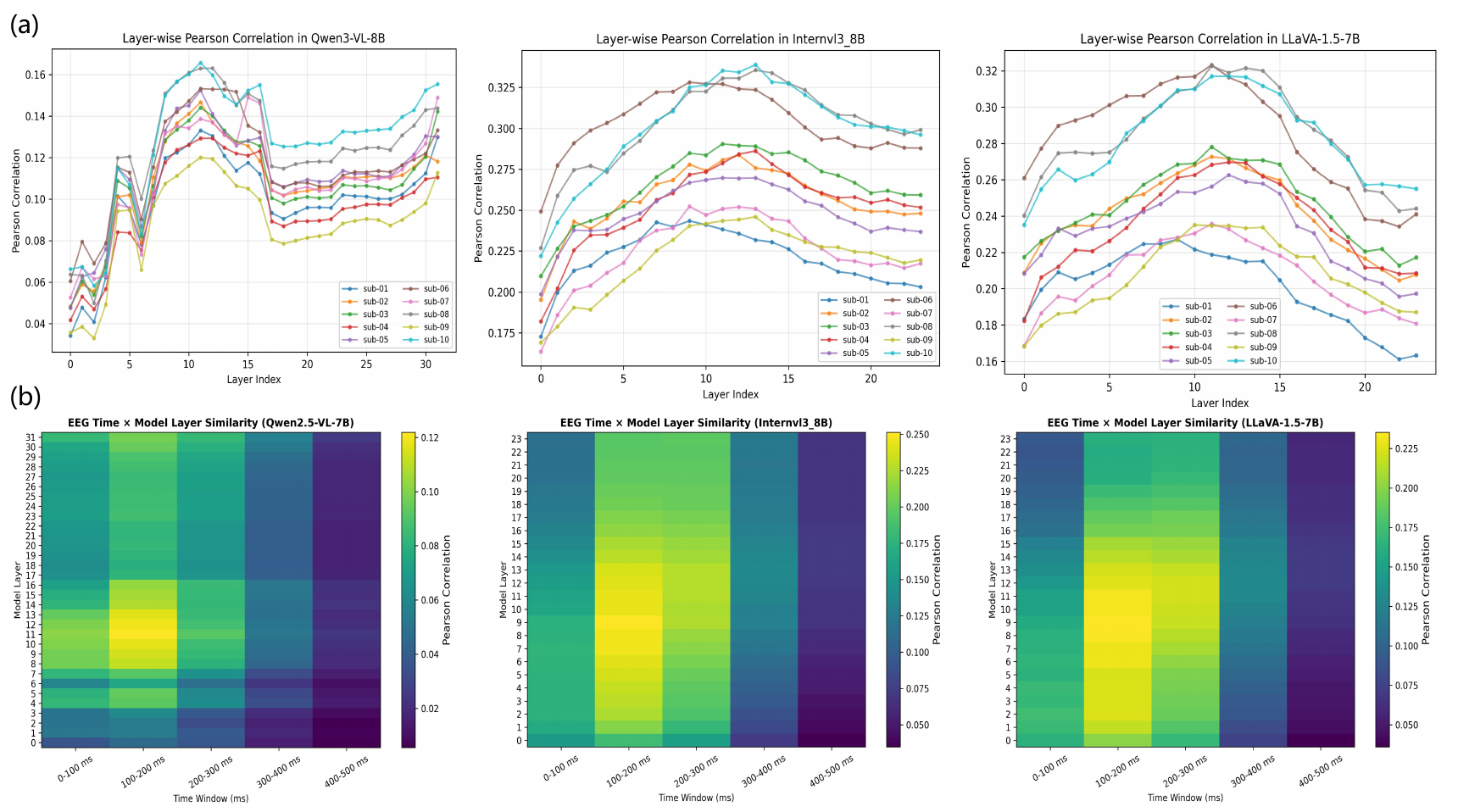}
  \caption{Layer-wise correspondence between the LVLMs and EEG signals across multiple participants.}
  \label{fig:layer}
\end{figure*}

% \paragraph{Statistical Significance Test}
% To assess the significance of the model’s predictions, we conducted a paired t-test comparing the Pearson correlation between true and predicted EEG signals against a distribution obtained from 200 shuffled EEG sets. A p-value below 0.05 indicated that the model’s performance exceeded chance.

\section{Experiments}

\subsection{Implementation Details} 
% All experiments are implemented in Python 3.10.18 with CUDA 12.2 and PyTorch 2.8.0. 
Experiments are conducted on the THINGS-EEG dataset \cite{gifford2022large} with a single NVIDIA A800 GPU. Statistical significance tests and additional experimental details are provided in Appendix~\ref{sec:Statistical}.
We used 32 LVLMs from 9 different model families, covering both vision-only and vision--language architectures with varying model scales, including ViT-based models \cite{dosovitskiy2020image}, Qwen2.5-VL \cite{bai2025qwen2}, Qwen3-VL \cite{bai2025qwen3vltechnicalreport}, LLaVA-v1.5 \cite{liu2024improved}, LLaVA-NeXT \cite{liu2024llavanext}, InternVL3 \cite{zhu2025internvl3exploringadvancedtraining}, InternVL3.5 \cite{wang2025internvl35advancingopensourcemultimodal}, DeepSeek-VL2 \cite{wu2024deepseekvl2mixtureofexpertsvisionlanguagemodels}, and SAIL-VL2 \cite{yin2025sailvl2technicalreport},
% including ViT-based models \cite{dosovitskiy2020image}, Qwen-VL series, LLaVA variants, InternVL, DeepSeek-VL, and SAIL-VL. 
All models employed in this study are publicly available from \url{https://huggingface.co/models}. More details are provided in Appendix~\ref{sec:LVLM}.

% We train the model for 50 epochs with a batch size of 1024, using a learning rate of $1\times10^{-4}$ for intra-subject and $1\times10^{-5}$ for inter-subject settings. Optimization is performed with AdamW \cite{loshchilov2018decoupled} and a weight decay of $1\times10^{-4}$. Early stopping is applied to prevent overfitting. To ensure stability, we use the Softplus function for the temperature parameter $\tau$.
% We use the visual branches of CLIP models with pretrained weights from OpenCLIP~\cite{ilharco2021openclip}. RN50 serves as the image encoder for retrieval, while ViT-H-14 is used for visual reconstruction.

\subsection{LVLM-brain Similarity Results}
\paragraph{Main Experimental Results.}
We conduct experiments following the method and evaluation metrics detailed in Section~\ref{method}. Table~\ref{tab:vlm_eeg_similarity} presents a comprehensive comparison of EEG alignment performance across 32 models. From the results, we could find that there is a highly consistent correlation among all the measurement indicators. In addition, all modern LVLMs exhibit statistically significant correlations with EEG signals, substantially exceeding random baselines; for example, Qwen2.5-VL-7B achieves 0.2316 ± 0.0280, whereas the random baseline remains close to zero at 0.0002 ± 0.0077. 
Cross-subject experiments are provided in Appendix~\ref{sec:subj}.
The RDM visualizations are provided in Appendix~\ref{sec:rdm}.
These results indicate that the high-level visual–semantic representations learned by VLMs capture meaningful neural information related to human visual processing. 

% leveraging multiple complementary evaluation metrics

% \section{RDM Alignment} 
% Using Qwen2.5-VL 7B as an example, we quantified the alignment between its visual representations and neural responses by computing representational dissimilarity matrices for both actual and model-predicted EEG signals (Figure~\ref{fig:rdm}). The strong correspondence indicates that Qwen2.5-VL captures key structural features of human neural representations during image processing.

% \begin{figure}[t]
%   \includegraphics[width=\columnwidth]{fig/rdm.pdf}
%   \caption{RDMs for predicted and actual EEG responses.}
%   \label{fig:rdm}
% \end{figure}

\paragraph{Differences Across Model Families.}
The models exhibit clear performance tiers in EEG alignment. The InternVL3.5 series achieves the highest performance, with its 38B model scoring Pearson 0.2649, Spearman 0.2573, CKA 0.4432, and RSA 0.3422, demonstrating the strongest overall alignment. The second tier is dominated by the Qwen series and LLaVA-Next models. In contrast, pure visual ViT models and LLaVA-v1.5 perform lowest, with ViT below 0.227 and LLaVA-v1.5 around 0.210, well behind mainstream LVLMs.
In particular, ViT-B, ViT-L, and ViT-H underperform most LVLMs, indicating that unimodal visual pretraining is insufficient to capture EEG-relevant neural representations. This gap highlights the importance of multimodal, language-augmented supervision for learning visual features that better align with human brain responses.

% The models exhibit clear performance tiers in EEG alignment. The InternVL3.5 series achieves the highest performance, with its 38B model scoring Pearson 0.2649, Spearman 0.2573, CKA 0.4432, and RSA 0.3422, demonstrating the strongest overall alignment. The InternVL3 family follows closely in the first tier. The second tier is dominated by the Qwen series and LLaVA-Next models, with Qwen3-VL 8B and LLaVA-Next 7B reaching the highest Pearson correlations at 0.2385 and 0.2377. In contrast, pure visual ViT models and LLaVA-v1.5 perform lowest, with ViT below 0.227 and LLaVA-v1.5 around 0.210, well behind mainstream LVLMs.

% \paragraph{Comparison Between Vision-only Models and LVLMs.}
% Pure vision models such as ViT-B, ViT-L, and ViT-H consistently underperform LVLMs in Pearson and Spearman correlation, indicating that unimodal visual pretraining is insufficient to capture EEG-relevant neural representations. 
% This contrast underscores the critical role of multimodal, language-augmented supervision in learning visual representations that more closely align with human brain responses.

% Pure vision models such as ViT-B, ViT-L, and ViT-H consistently exhibit lower Pearson and Spearman correlations than most LVLMs, suggesting that unimodal visual pretraining is insufficient to fully capture EEG-relevant neural representations. 

\begin{figure*}[t]
\centering
  \includegraphics[width=0.92\textwidth]{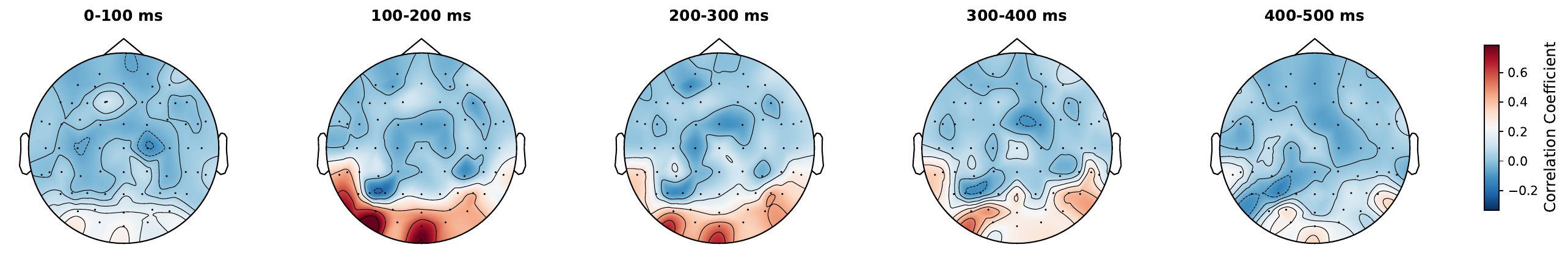}
  \caption{Spatiotemporal dynamics of LVLM-EEG similarity.}
  \label{fig:topo}
\end{figure*}

\begin{table*}[t]
\centering
\footnotesize
\setlength{\tabcolsep}{4.5pt}
\resizebox{0.85\textwidth}{!}{
\begin{tabular}{@{}p{3.8cm}ccccc@{}}
\toprule
\textbf{Input Configuration} & \textbf{Pearson} & \textbf{Spearman} & \textbf{CKA} & \textbf{RSA} & \textbf{Kendall} \\
\midrule

\multicolumn{6}{c}{\textit{Visual Encoder Only}} \\
\midrule
\rowcolor{gray!8}
Image only & 0.232 $\pm$ 0.028 & 0.225 $\pm$ 0.026 & \textbf{0.415 $\pm$ 0.097} & \textbf{0.298 $\pm$ 0.088} & 0.154 $\pm$ 0.018 \\

\midrule
\multicolumn{6}{c}{\textit{Multimodal Fusion}} \\
\midrule

\rowcolor{blue!8}
Image + Explicit prompt & \textbf{0.233 $\pm$ 0.032} & \textbf{0.226 $\pm$ 0.030} & 0.406 $\pm$ 0.097 & 0.289 $\pm$ 0.092 & \textbf{0.197 $\pm$ 0.064} \\
\rowcolor{blue!8}
Image + No prompt & 0.233 $\pm$ 0.032 & 0.226 $\pm$ 0.030 & 0.406 $\pm$ 0.097 & 0.289 $\pm$ 0.092 & 0.197 $\pm$ 0.064 \\
\rowcolor{blue!8}
Image + Noise prompt & 0.233 $\pm$ 0.032 & 0.226 $\pm$ 0.030 & 0.406 $\pm$ 0.097 & 0.289 $\pm$ 0.092 & 0.197 $\pm$ 0.064 \\
\rowcolor{blue!8}
Image + Caption prompt & 0.233 $\pm$ 0.032 & 0.226 $\pm$ 0.030 & 0.406 $\pm$ 0.097 & 0.289 $\pm$ 0.092 & 0.197 $\pm$ 0.064 \\
\rowcolor{red!8}
\textit{Black image + Caption prompt} & \textit{0.171 $\pm$ 0.029} & \textit{0.168 $\pm$ 0.029} & \textit{0.311 $\pm$ 0.069} & \textit{0.177 $\pm$ 0.067} & \textit{0.119 $\pm$ 0.045} \\
\bottomrule
\end{tabular}
}
\caption{Representation analysis of QwenVL2.5 under different input configurations.}
\label{tab:qwenvl_modality_analysis}
\end{table*}

\paragraph{Impact of Model Scale.} 
While some models show modest gains with increasing parameter size, larger models do not consistently perform better across families. For instance, in the Qwen3-VL series, the 8B model outperforms the 32B model in Pearson correlation, and within InternVL3.5, scaling from 2B to 38B yields only a 1.6 percentage point improvement. In contrast, architectural design has a much larger impact: InternVL3.5-38B achieves a Pearson correlation of 0.2649, compared to 0.2102 for LLaVA-v1.5-7B, a gap more than three times larger than the maximum gain from scaling alone. These results indicate that architectural choices, rather than model size, are the primary driver of improved alignment with neural signals, consistent with prior findings on cortex-aligned representations \cite{kazemian2025convolutional}.

\subsection{Consistency with Human Visual}

\paragraph{Hierarchical LVLM-to-EEG Temporal Alignment.}
The layer-wise correspondence between LVLM  representations and EEG signals exhibits a clear hierarchical and temporal structure. As shown in Figure~\ref{fig:layer}~(a), Pearson correlations consistently peak at intermediate model layers across subjects, indicating the strongest and most stable neural alignment. Temporal analysis in Figure~\ref{fig:layer}~(b) further shows that these intermediate layers, roughly layers 8–16, align most strongly with EEG activity in the 100–300 ms window, while shallow and deep layers, as well as early and late time intervals, show markedly weaker correspondence. These results suggest that intermediate LVLM representations temporally align with high-level visual cognitive processing, consistent with the hierarchical organization of the visual cortex \cite{babiloni2006visuo} and recent evidence for hierarchical brain recapitulation in deep models \cite{muttenthaler2025aligning}.

% The layer-wise correspondence between LVLM visual representations and EEG signals reveals a clear hierarchical and temporal alignment. As shown in Figure~\ref{fig:layer}~(a), Pearson correlations consistently peak in intermediate model layers across subjects, indicating that these representations exhibit the strongest and most stable neural alignment. This pattern is further supported by the temporal analysis in Figure~\ref{fig:layer}~(b), which maps LVLM layers against EEG time windows (0–500 ms). Intermediate layers (approximately layers 8–16) show the strongest associations with EEG activity in the 100–300 ms window, whereas alignment is substantially weaker for shallow and deep layers as well as during early (0–100 ms) and late (300–500 ms) intervals. Together, these results suggest that intermediate LVLM representations align temporally with high-level visual cognitive processing stages reflected in human EEG.
% This align with early visual EEG stages and later layers with higher-level cognitive processing, echoing visual cortex organization \cite{babiloni2006visuo} and recent findings on hierarchical brain recapitulation \cite{muttenthaler2025aligning}.

\paragraph{Spatiotemporal Dynamics of Similarity.}
Figure~\ref{fig:topo} illustrates the spatiotemporal dynamics of EEG–LVLM image feature correspondence within the first 500 ms, using 100 ms time windows. During the 0–100 ms interval, correlations are generally low across the brain. Between 100–300 ms, strong correlations emerge in the occipital region and gradually spread, peaking in intensity. From 300–400 ms, high-correlation areas extend toward the parietal cortex, followed by a decline to lower levels in the 400–500 ms window. These patterns reflect the temporal progression of visual processing, with early occipital engagement followed by parietal propagation, consistent with the spatiotemporal dynamics of human visual information processing \cite{robinson2017very}.
Results by brain region are in Appendix~\ref{sec:Regions}.

% Results for different brain regions are provided in Appendix~\ref{sec:Regions}.

% Figure~\ref{fig:topo} illustrates the spatiotemporal dynamics of EEG–LVLM image feature correspondence within the first 500 milliseconds, using 100 ms time windows. During the 0–100 ms interval, correlations are generally low across the brain. Between 100–300 ms, strong correlations emerge in the occipital region and gradually spread, peaking in intensity. From 300–400 ms, high-correlation areas extend toward the parietal cortex, followed by a decline to lower levels in the 400–500 ms window. These patterns reflect the temporal progression of visual processing, with early occipital engagement followed by parietal propagation, consistent with the spatiotemporal dynamics of human visual information processing.

\begin{figure}[t]
 \centering
  \includegraphics[width=0.9\columnwidth]{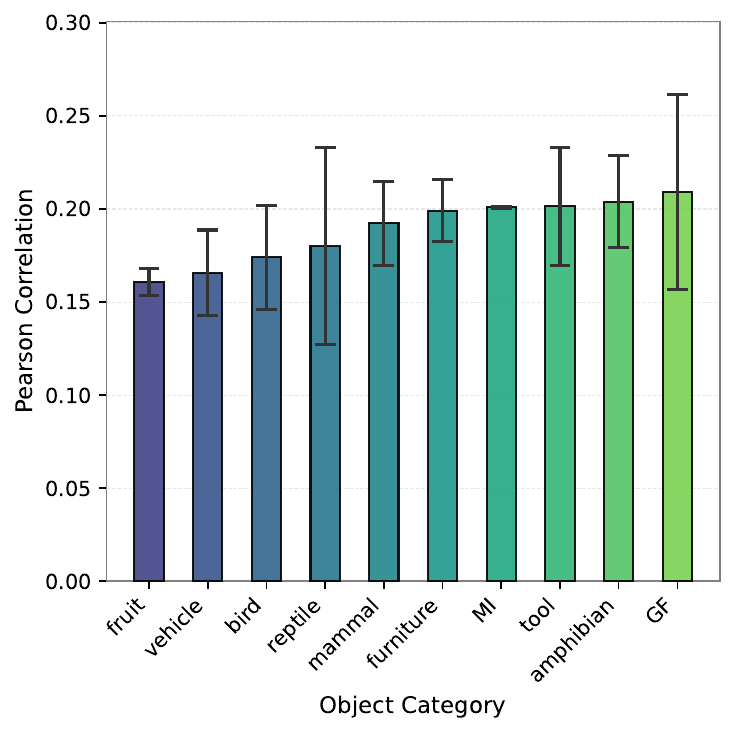}
  \caption{Category-level correlations, where MI is musical instrument and GF is geological formation.}
  \label{fig:category}
\end{figure}

\begin{figure*}[t]
  \includegraphics[width=\textwidth]{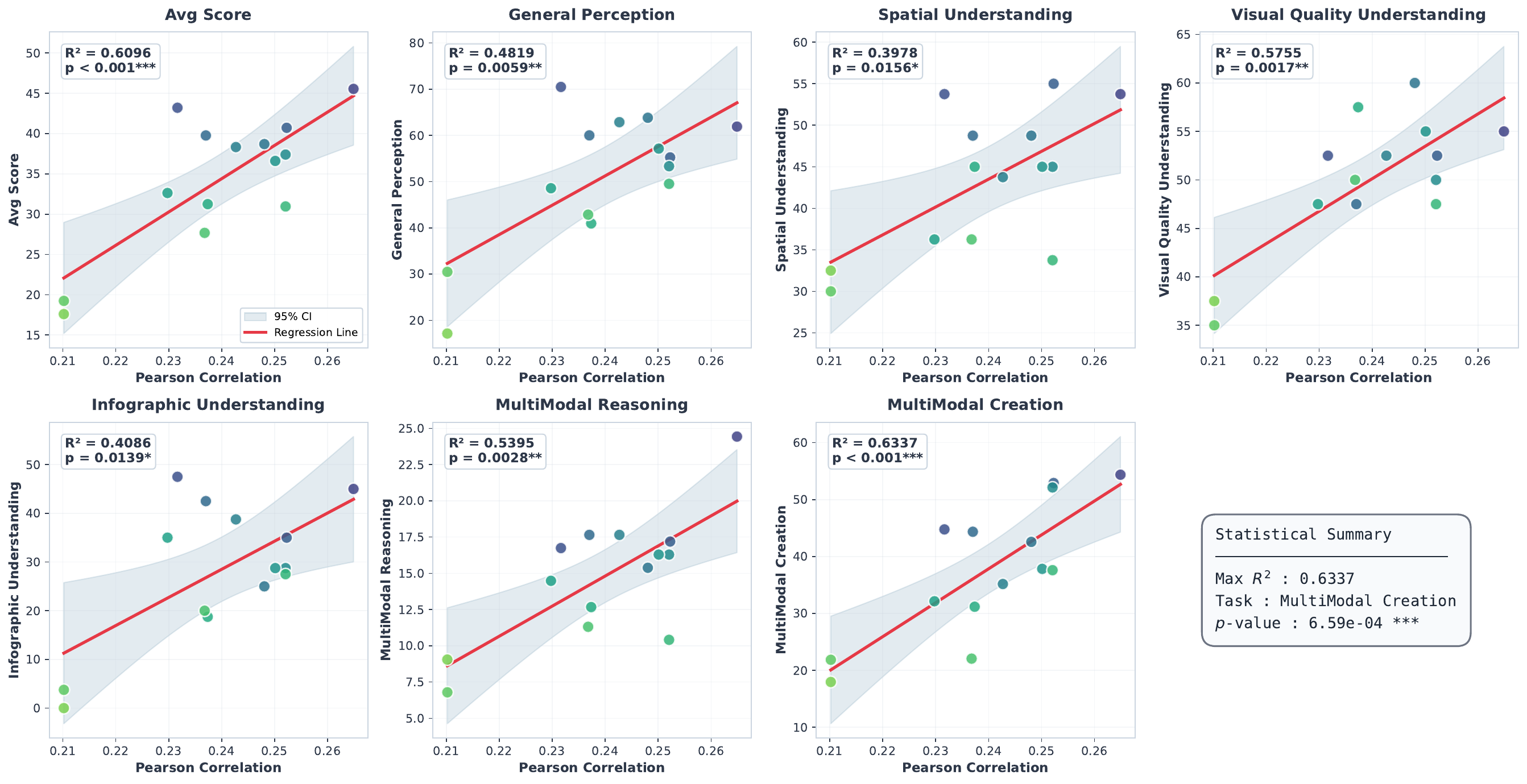}
  % \caption{Relationship between LVLM benchmark performance and LVLM-brain similarity.}
  \caption{Relationship between LVLM benchmark performance and LVLM-brain similarity. The red line denotes the regression line of the relationship, while the shaded area represents the 95\% confidence interval of the regression.}
  \label{fig:performance}
\end{figure*}

\subsection{Cross-Modal Layer Consistency}
Table~\ref{tab:qwenvl_modality_analysis} shows that multimodal fusion in QwenVL2.5 consistently improves rank-based alignment over the visual-only setting, yielding the highest Pearson, Spearman, and Kendall correlations. Performance is nearly identical across prompt types—including an explicit prompt (“Describe this image.”), empty prompt, noise prompt with five random words (e.g., “Harmony illuminate umbrella freedom like.”), and caption prompt, indicating that gains arise from multimodal integration rather than prompt semantics. In contrast, the \emph{black image + caption} condition degrades all metrics, confirming the necessity of valid visual input. Notably, the visual-only encoder achieves the highest CKA and RSA, suggesting stronger linear and geometric structure that is partially smoothed by cross-modal fusion.

\subsection{Category-Level Similarity}
% Figure~\ref{fig:category} shows correlations between LVLM representations and EEG signals across object categories, revealing clear category-dependent differences. All categories exhibit correlations above 0.15, with geological formation (GF) and amphibian reaching values around 0.21, while fruit shows the lowest similarity at approximately 0.15, and the small error bars indicate stable measurements. These differences likely reflect both human visual processing and model training biases, as animal-related categories evoke richer and more discriminative neural representations and are associated with more diverse visual–semantic information during LVLM pretraining, leading to stronger alignment, whereas visually and semantically simpler categories such as fruit show weaker correspondence.

Figure~\ref{fig:category} shows category-specific similarities between LVLM representations and EEG signals, revealing clear differences across object categories. All categories exceed chance-level alignment, but alignment strength varies: geological formations and amphibians achieve the highest correlations, around 0.21, while vehicles and fruits show the weakest similarity, approximately 0.15. The narrow error bars confirm stable measurements.
These discrepancies reflect how neural and model representations interact. Categories such as amphibians evoke richer and more discriminative neural patterns and are also associated with diverse visual-semantic information in model training, leading to stronger alignment. In contrast, visually and semantically simpler categories like fruits exhibit weaker correspondence, suggesting their neural representations may rely more on low-level perceptual features that are less emphasized in high-level LVLM embeddings. Thus, neural alignment depends jointly on the biological salience of visual categories and the semantic hierarchies learned by the model.

% \subsection{Correlation with LVLM Benchmark Evaluations}
\subsection{Correlation with LVLM Benchmarks}
Figure~\ref{fig:performance} shows the relationship between LVLM–brain similarity and benchmark performance on seven tasks from OpenCompass (\url{https://opencompass.org.cn/}), including average score, general perception, spatial understanding, visual quality understanding, infographic understanding, multimodal reasoning, and multimodal creation, more details are provided in Appendix~\ref{sec:OpenCompass}. Across all subplots, the regression lines display clear positive trends, indicating that higher Pearson similarity with brain signals is associated with better task performance. This relationship is strongest for the overall average score ($R^2=0.6096$, $p<0.001$), highlighting a robust link between neural alignment and model capability. The narrow 95\% confidence intervals further support the statistical stability of this trend. Task-level analysis reveals that multimodal creation ($R^2=0.6337$, $p<0.001$) and multimodal reasoning ($R^2=0.5395$, $p=0.0028$) exhibit the strongest associations, whereas spatial understanding shows the weakest and least significant correlation ($R^2=0.3978$, $p=0.0156$), suggesting a weaker coupling between neural similarity and spatial task performance.

\section{Conclusions}
To investigate the neural alignment of LVLMs, we applied ridge regression and RSA to image-evoked EEG. Our results show that LVLMs align significantly with human brain, with intermediate layers coupling strongest to mid-stage activity. Model architectural design outweighs parameter scaling in achieving brain-aligned representations, and LVLMs outperform unimodal vision models. Alignment peaks in occipital–parietal regions at 100–300ms, matching human visual processing dynamics. Higher neural alignment correlates with better LVLM benchmark performance, revealing clear correspondences between artificial and biological visual processing and offer insights for brain-inspired multimodal model design.

\section*{Limitations}
This study is based on visual representations of open-source LVLMs, precluding the evaluation of LVLM-brain similarity for closed-source LVLMs (e.g., GPT-4V). We aim to extend these findings to more LVLM variants in subsequent research. Additionally, EEG has limited spatial resolution for brain activity, cannot achieve single-neuron recording and poorly captures deep neural signal sources, meaning the precise neural substrates of our findings require further exploration with intracranial recordings. Moreover, constrained by available aligned EEG-image datasets, our research was performed with a relatively limited set of visual stimuli.

\section*{Ethics Statement}
This study analyzes previously collected, publicly available EEG datasets and open-source LVLM checkpoints. We do not collect new human-subject data, and we do not attempt to identify, re-contact, or infer personal attributes of any participant. We therefore believe that this study complies with the ACL Ethics Policy.

\bibliography{custom}

\appendix

\section{Appendix}
% \label{sec:appendix}

\subsection{Implementation Details}
\label{sec:Statistical}
All the experiments were implemented in Python 3.10.18 with CUDA 12.2 and PyTorch 2.8.0. Experiments are conducted on a single NVIDIA A800 GPU. For each subject, we perform PCA to reduce EEG signals to 256 dimensions (explaining >80\% variance) and apply standardization to both EEG and visual features. We train a ridge regression model with 5-fold cross-validation to select the optimal regularization parameter ($\alpha \in [10^{-2}, 10^{3}]$). 
To assess the significance of the model’s predictions, we conducted a paired t-test comparing the Pearson correlation between true and predicted EEG signals against a distribution obtained from 200 shuffled EEG sets. A p-value below 0.05 indicated that the model’s performance exceeded chance.

\subsection{Details on the LVLMs}
\label{sec:LVLM}
We used 32 LVLMs drawn from 9 LVLM Series. Below, we provide a detailed description of each model series.

\paragraph{ViT Series (Vision Transformer)}
The Vision Transformer (ViT) series, revolutionized computer vision by adapting the Transformer architecture from natural language processing to image data. Unlike convolutional neural networks (CNNs) that rely on local receptive fields, ViT splits images into fixed-size patches, embeds them as tokens, and processes them with multi-head self-attention mechanisms. The ViT-B (Base), ViT-L (Large), and ViT-H (Huge) variants differ in model depth, number of attention heads, and hidden dimension size—scaling up these parameters consistently improves performance on large-scale datasets such as ImageNet-21k and JFT-300M. ViT established a new paradigm for vision tasks, laying the foundation for subsequent vision-language models (VLMs) by demonstrating the efficacy of Transformer-based architectures in capturing global image context.
\paragraph{Qwen2.5-VL Series}
The Qwen2.5-VL series, developed by Alibaba Cloud, is a suite of open-source vision-language models presented in technical reports associated with the Qwen model family, building on the advancements of Qwen2-VL. Comprising Qwen2.5-VL-3B/7B/32B/72B, these models integrate a frozen vision encoder (based on an improved ViT variant) with a decoder-only language model, optimized for multi-modal tasks including image captioning, visual question answering (VQA), and image-grounded dialogue. Key enhancements over previous iterations include refined cross-modal alignment strategies, better handling of high-resolution images, and improved robustness to noisy visual inputs. The series supports diverse input formats (e.g., text, images, multi-image sequences) and achieves state-of-the-art results on benchmarks like MME, SEED-Bench, and VQAv2, with larger models (32B/72B) excelling in complex reasoning tasks while smaller variants (3B/7B) offer efficient deployment for edge devices.
\paragraph{Qwen3-VL Series}
The Qwen3-VL series represents the latest vision-language iteration of Alibaba Cloud’s Qwen model family, introduced in 2024 with technical documentation highlighting advancements in cross-modal fusion and efficiency. Featuring four variants (2B/4B/8B/32B), Qwen3-VL adopts a unified encoder-decoder architecture with a lightweight yet powerful vision encoder and a language decoder pre-trained on a massive corpus of image-text pairs, web documents, and multi-modal dialogues. Compared to Qwen2.5-VL, it incorporates dynamic patch embedding to adapt to varying image resolutions and a novel cross-attention mechanism that enhances alignment between visual features and textual tokens. Qwen3-VL demonstrates superior performance on low-resource visual reasoning tasks and supports real-world applications such as document understanding, medical image analysis, and multi-turn visual dialogue, with the 32B model leading in academic benchmarks and smaller models (2B/4B) optimized for latency-sensitive scenarios.
\paragraph{LLaVA-v1.5 Series}
The LLaVA-v1.5 series is a lightweight yet effective vision-language model built by aligning a CLIP ViT-L/14 vision encoder with LLaMA-2-based language decoders (7B/13B). Unlike its predecessor LLaVA-v1, LLaVA-v1.5 leverages a refined visual instruction tuning dataset (LLaVA-Instruct-150K) and optimized cross-modal projection layers to bridge the gap between visual and textual representations. The model excels in open-ended visual question answering, image captioning, and visual grounding tasks, achieving competitive results on benchmarks like GQA, COCO Caption, and VQAv2 while maintaining efficient inference speeds. Its open-source nature and compatibility with popular LLM frameworks have made LLaVA-v1.5 a widely adopted baseline for research on visual instruction tuning and lightweight VLM development.
\paragraph{LLaVA-Next Series}
The LLaVA-Next series, an evolution of the LLaVA model family presented in technical reports by the original LLaVA team, enhances cross-modal capabilities by integrating advanced vision encoders and updated language backbones (7B/13B variants based on LLaMA-2 and Mistral). Key improvements include a two-stage pre-training pipeline: first, aligning visual features with language models using image-text pairs, then fine-tuning on a diverse set of visual instruction data (including complex reasoning, multi-image tasks, and spatial understanding). LLaVA-Next introduces a dynamic visual context window to handle high-resolution images and multi-image inputs, enabling applications such as video frame analysis and multi-document visual reasoning. It outperforms LLaVA-v1.5 on benchmarks like MME and SEED-Bench, particularly in tasks requiring spatial and logical reasoning, while retaining the lightweight and deployable characteristics of the LLaVA family.
\paragraph{InternVL3 Series}
The InternVL3 series, developed by the Shanghai AI Laboratory, is a scalable vision-language model. Boasting six variants (1B/2B/8B/9B/14B/38B), InternVL3 adopts a hybrid architecture that combines a hierarchical vision encoder (supporting 4K-resolution images) with a multi-modal language decoder, pre-trained on over 1 billion high-quality image-text pairs. A core innovation is the spatial-aware cross-attention mechanism, which preserves fine-grained visual details (e.g., small objects, text in images) critical for tasks like document OCR, medical imaging, and remote sensing. InternVL3 sets new records on benchmarks such as DocVQA, TextVQA, and MME, with the 38B model delivering state-of-the-art performance in high-resolution visual reasoning and smaller variants optimized for edge and cloud deployment.

\begin{table*}[t]
\centering
\footnotesize
\setlength{\tabcolsep}{4.2pt}
\resizebox{\textwidth}{!}{
\begin{tabular}{@{}lccccccccc@{}}
\toprule
\textbf{Model} & \textbf{Language Model} & \textbf{Vision Model} & \textbf{Avg} & \textbf{Gen. Percep.} & \textbf{Spatial} & \textbf{Visual Quality} & \textbf{Infographic} & \textbf{MM Reason.} & \textbf{MM Creation} \\
\midrule

InternVL3-38B & Qwen2.5-32B & InternViT-6B-v2.5 & \textbf{45.55} & 61.90 & 53.75 & 55.00 & 45.00 & 24.43 & 54.37 \\
Qwen2.5-VL-7B & Qwen2.5-7B & QwenViT & 43.21 & \textbf{70.48} & 53.75 & 52.50 & \textbf{47.50} & 16.74 & 44.76 \\
InternVL3-14B & Qwen2.5-14B & InternViT-300M-v2.5 & 40.72 & 55.24 & \textbf{55.00} & 52.50 & 35.00 & 17.19 & 52.91 \\
SAIL-VL2-8B & Qwen3-8B & SAILViT-Huge & 39.77 & 60.00 & 48.75 & 47.50 & 42.50 & 17.65 & 44.35 \\
DeepSeek-VL2 & DeepSeekMoE-27B & SigLIP-400M & 38.70 & 63.81 & 48.75 & \textbf{60.00} & 25.00 & 15.38 & 42.57 \\
Qwen2.5-VL-3B & Qwen2.5-3B & QwenViT & 38.33 & 62.86 & 43.75 & 52.50 & 38.75 & 17.65 & 35.17 \\
InternVL3-8B & Qwen2.5-7B & InternViT-300M-v2.5 & 37.40 & 53.33 & 45.00 & 50.00 & 28.75 & 16.29 & 52.13 \\
DeepSeek-VL2-Small & DeepSeekMoE-16B & SigLIP-400M & 36.61 & 57.14 & 45.00 & 55.00 & 28.75 & 16.29 & 37.83 \\
SAIL-VL2-2B & Qwen3-1.7B & SAILViT-Huge & 32.63 & 48.57 & 36.25 & 47.50 & 35.00 & 14.48 & 32.14 \\
InternVL3-2B & Qwen2.5-1.5B & InternViT-300M-v2.5 & 30.96 & 49.52 & 33.75 & 47.50 & 27.50 & 10.41 & 37.60 \\
DeepSeek-VL2-Tiny & DeepSeekMoE-3B & SigLIP-400M & 27.68 & 42.86 & 36.25 & 50.00 & 20.00 & 11.31 & 22.06 \\
InternVL3-1B & Qwen2.5-0.5B & InternViT-300M-v2.5 & 24.39 & 37.14 & 35.00 & 40.00 & 15.00 & 9.50 & 24.58 \\
InternLM-XComposer2 & InternLM2-7B & CLIP ViT-L/14 & 31.25 & 40.95 & 45.00 & 57.50 & 18.75 & 12.67 & 31.18 \\
LLaVA-1.5-13B & Vicuna-v1.5-13B & CLIP ViT-L/14 & 19.24 & 30.48 & 30.00 & 35.00 & 3.75 & 6.79 & 21.86 \\
LLaVA-1.5-7B & Vicuna-v1.5-7B & CLIP ViT-L/14 & 17.60 & 17.14 & 32.50 & 37.50 & 0.00 & 9.05 & 17.96 \\

\bottomrule
\end{tabular}
}
\caption{Comparison of representative vision--language models on multimodal perception, reasoning, and creation benchmarks.}
\label{tab:vlm_benchmark_comparison}
\end{table*}

\paragraph{InternVL3.5 Series}
The InternVL3.5 series, the latest update to the InternVL family from the Shanghai AI Laboratory, builds on InternVL3 with refined model architectures and expanded pre-training data, as outlined in 2024 technical reports. The six variants (1B/2B/4B/8B/14B/38B) introduce a unified vision-language tokenizer that merges visual patches and textual tokens into a shared embedding space, reducing cross-modal alignment loss and improving inference efficiency. Key enhancements include support for multi-image sequences (e.g., video clips, panoramic images) and a lightweight adapter module that enables fine-tuning on domain-specific tasks (e.g., medical imaging, autonomous driving) without retraining the entire model. InternVL3.5 outperforms InternVL3 on both general and specialized benchmarks, with the 4B variant offering a balanced trade-off between performance and latency, making it suitable for real-world multi-modal applications.
\paragraph{ DeepSeek-VL2 Series}
The DeepSeek-VL2 series, developed by DeepSeek AI, is a compact yet powerful vision-language model suite consisting of Tiny, Small, and Base variants. DeepSeek-VL2 adopts a distilled vision encoder (based on MobileViT) paired with a lightweight language decoder, optimized for low-latency inference on edge devices and mobile platforms. A key feature is the multi-scale visual feature fusion module, which extracts both local and global image features to handle diverse tasks (e.g., VQA, image captioning, object detection) with minimal computational overhead. The series achieves competitive performance on lightweight VLM benchmarks while maintaining fast inference speeds—making it ideal for embedded systems, mobile apps, and real-time visual interaction scenarios.
\paragraph{SAIL-VL2 Series}
The SAIL-VL2 series, developed by the SAIL (Shenzhen Academy of Artificial Intelligence) team, is a pair of efficient vision-language models (2B/8B) designed for edge and cloud deployment, as presented in 2024 technical publications. SAIL-VL2 integrates a lightweight vision transformer (ViT-Small) with a decoder-only language model pre-trained on a curated dataset of Chinese and English image-text pairs, emphasizing cross-lingual visual reasoning capabilities. Unique to SAIL-VL2 is the adaptive cross-modal adapter, which dynamically adjusts the strength of visual-textual fusion based on task complexity, improving performance on both simple (e.g., image classification) and complex (e.g., visual commonsense reasoning) tasks. The 2B variant is optimized for edge devices with limited computational resources, while the 8B variant delivers enhanced performance on academic benchmarks like VQAv2 and GQA, making SAIL-VL2 suitable for both research and industrial applications.

\begin{figure*}[t]
  \includegraphics[width=\textwidth]{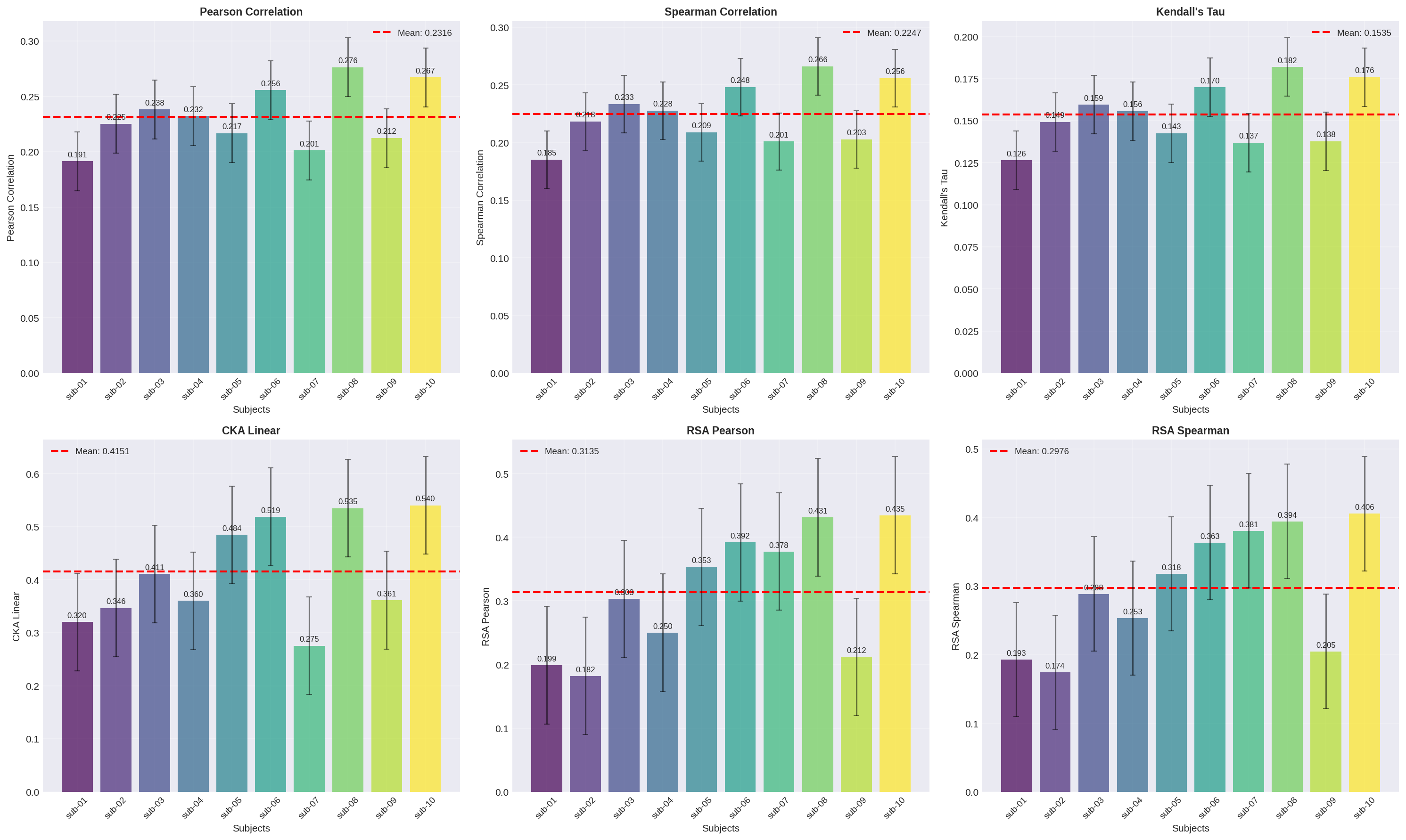}
  \caption{Quantifying Similarity Between LVLM Representations and EEG Signals Across Subjects.}
  \label{fig:subj}
\end{figure*}

\begin{figure}[t]
  \includegraphics[width=\columnwidth]{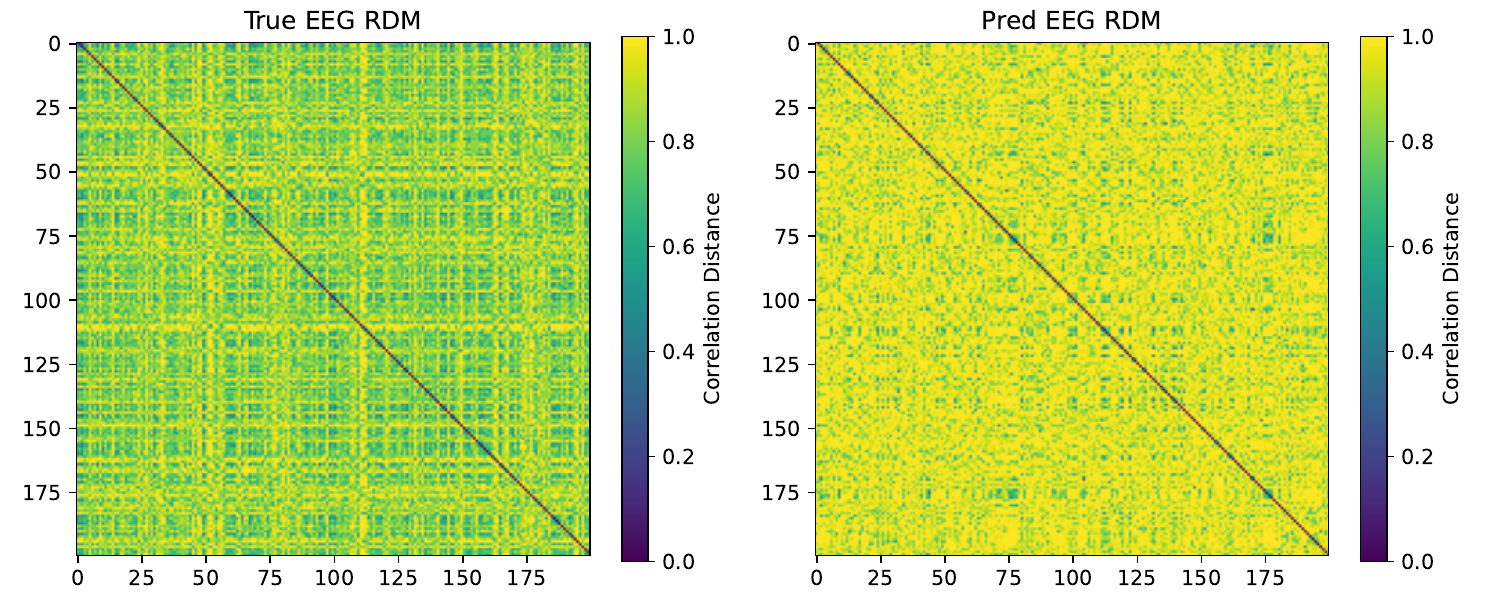}
  \caption{RDMs for predicted and actual EEG responses.}
  \label{fig:rdm}
\end{figure}

\begin{figure}[t]
  \includegraphics[width=\columnwidth]{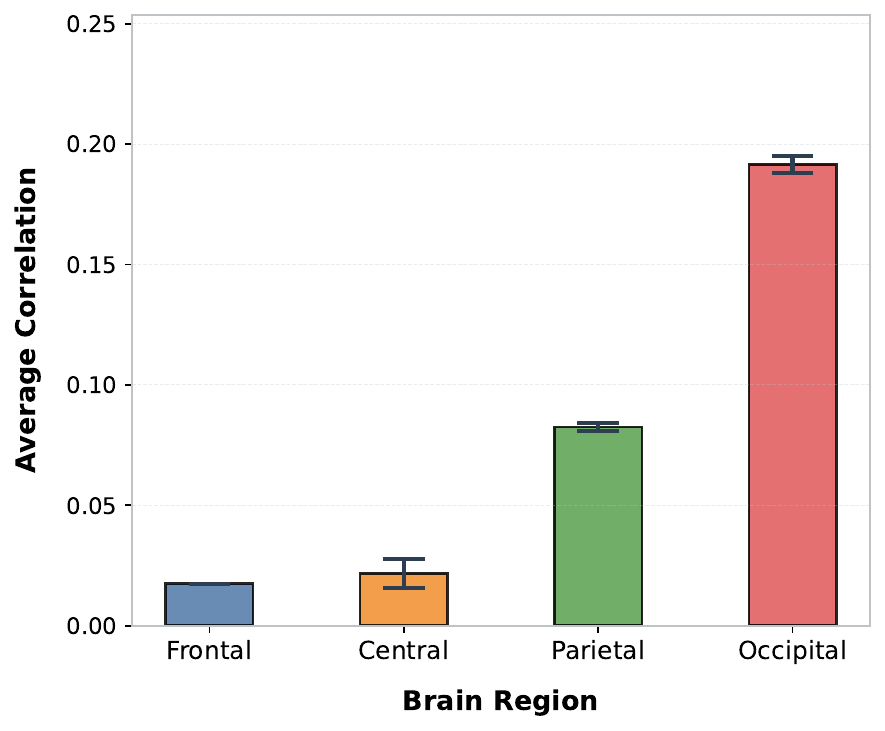}
    \caption{Mean correlations between EEG signals from frontal, central, parietal, and occipital brain regions and LVLM representations.}
  \label{fig:region}
\end{figure}
% 这个图要美化

\subsection{OpenCompass Multi-modal Leaderboard}
\label{sec:OpenCompass}
We further report model performance based on the \textbf{OpenCompass Multi-modal Leaderboard}, with results obtained from the official real-time leaderboard at
\url{https://rank.opencompass.org.cn/leaderboard-multimodal-official/?m=REALTIME}.

The leaderboard provides a systematic evaluation of state-of-the-art multimodal models across multiple capability dimensions. Models are ranked in descending order according to a \emph{weighted average score}, computed from their performance across all evaluated dimensions.

Specifically, the current OpenCompass multi-modal benchmark assesses models along the following six capability dimensions:
\begin{itemize}
    \item \textbf{General Perception}: the ability to recognize and understand entities, events, and complex visual--textual content in natural scenes.
    \item \textbf{Spatial Awareness}: the capability to perceive and reason about both 2D planar layouts and 3D spatial structures.
    \item \textbf{Visual Quality Assessment}: the capacity to evaluate image quality attributes, aesthetic properties, and subjective visual effects.
    \item \textbf{Infographic Understanding}: the ability to analyze structured or artificially designed visual content such as charts and diagrams.
    \item \textbf{Multimodal Reasoning}: the ability to solve mathematical or scientific problems spanning multiple modalities and to perform complex logical reasoning.
    \item \textbf{Multimodal Creation}: the capability to understand cross-modal instructions and generate creative content in both social and professional contexts.
\end{itemize}

Given the importance of reasoning in multimodal intelligence, the \textbf{Multimodal Reasoning} dimension is assigned \emph{twice the weight} of the other dimensions when computing the overall weighted average score.

\subsection{RDM Alignment} 
\label{sec:rdm}
Using Qwen2.5-VL 7B as an example, we quantified the alignment between its visual representations and neural responses by computing representational dissimilarity matrices for both actual and model-predicted EEG signals (Figure~\ref{fig:rdm}). The strong correspondence indicates that Qwen2.5-VL captures key structural features of human neural representations during image processing.

\subsection{LVLM-EEG Similarity Across Subjects}
\label{sec:subj}
Figure~\ref{fig:subj} characterizes LVLM–EEG similarity across 10 subjects via \textsc{Qwen2.5-VL-7B}. Substantial inter-subject variability emerges: subjects such as \texttt{subj08} demonstrate uniformly elevated alignment, while counterparts like \texttt{subj01} exhibit attenuated values. Across all metrics, the mean similarity (denoted by the red dashed line) peaks for linear CKA at 0.4151 and reaches a minimum for Kendall’s Tau at 0.1355. These results reveal that the magnitude of LVLM–EEG alignment is highly sensitive to the choice of similarity metric. Critically, however, consistent LVLM–EEG similarity is observed across all subjects, regardless of metric, confirming the robustness of the observed alignment to individual differences in neural or behavioral profiles.

% Figure~\ref{fig:subj} shows the LVLM–EEG similarity across 10 subjects using \textsc{Qwen2.5-VL-7B}. Similarity values vary across subjects, with some individuals (e.g., \texttt{subj08}) exhibiting consistently higher alignment and others (e.g., \texttt{subj01}) showing lower values. Across metrics, the mean similarity (red dashed line) is highest for linear CKA (0.4151) and lowest for Kendall’s Tau (0.1355), indicating that while the magnitude of alignment is strongly metric-dependent, LVLM–EEG similarity is consistently observed across subjects.

% Figure~\ref{fig:subj} quantifies the similarity between LVLM representations and EEG signals across 10 subjects. Across metrics, similarity values vary by subject (e.g., subject subj08 often shows higher values, while subj01 is relatively lower), and the mean similarity (red dashed line) is highest for CKA Linear (0.4151) and lowest for Kendall’s Tau (0.1355), indicating that LVLM-EEG similarity is metric-dependent but consistently present across most subjects.

\subsection{Different Brain Regions} 
\label{sec:Regions}
Figure~\ref{fig:region}  shows the mean correlations between EEG signals from different brain regions, including the frontal, central, parietal, and occipital regions, and VLM representations. Clear regional differences are observed, with the strongest alignment in the occipital region, followed by the parietal region, while the frontal and central regions exhibit very weak correlations. This suggests that LVLM–EEG correspondence is most pronounced in visually related brain areas and substantially weaker in other regions.

\end{document}